\documentclass[12pt]{article}

\usepackage{amsmath}
\usepackage{amssymb}
\usepackage{cite}
\usepackage{theorem}
\usepackage{myLatexmacros}
\usepackage{graphicx}
\usepackage{enumerate}
\usepackage{color}
\definecolor{red}{rgb}{1, 0, 0}

\newtheorem{theorem}{\textbf{Theorem}}
\newtheorem{lemma}{\textbf{Lemma}}

\newtheorem{corollary}{\textbf{Corollary}}
\newtheorem{fact}{\textbf{Fact}}

\begin{document}

\title{Capacity-Achieving Codes with Bounded Graphical Complexity on Noisy Channels}

\author{\normalsize Chun-Hao Hsu and Achilleas Anastasopoulos\\
    \small Electrical Engineering and Computer Science Department\\
    \small University of Michigan\\
    \small Ann Arbor, MI, 48109-2122\\
    \small email: \{chhsu, anastas\}@umich.edu
}
\date{}

\maketitle

\thispagestyle{empty}

\begin{abstract}

We introduce a new family of concatenated codes with an outer
low-density parity-check (LDPC) code and an inner low-density
generator matrix (LDGM) code, and prove that these codes can
achieve capacity under any memoryless binary-input
output-symmetric (MBIOS) channel using maximum-likelihood (ML)
decoding with bounded graphical complexity, \ie, the number of
edges per information bit in their graphical representation is
bounded.
We also show that these codes can achieve capacity for the special
case of the binary erasure channel (BEC) under belief propagation
(BP) decoding with bounded decoding complexity per information bit
for all erasure probabilities in (0, 1).
By deriving and analyzing the average weight distribution (AWD)
and the corresponding asymptotic growth rate of these codes with a
rate-1 inner LDGM code, we also show that these codes achieve the
Gilbert-Varshamov bound with asymptotically high probability.
This result can be attributed to the presence of the inner rate-1
LDGM code, which is demonstrated to help eliminate high weight
codewords in the LDPC code while maintaining a vanishingly small
amount of low weight codewords.

\end{abstract}

\normalsize
\section{Introduction}
\label{intro}

During the last decade, several codes have been found to achieve
capacity on the binary erasure channel (BEC) under iterative
decoding. The first well-known example is the low-density
parity-check (LDPC) codes, which were introduced by
Gallager~\cite{Ga62} and proved to be capacity-achieving about
forty years later~\cite{Sh99,OsSh02}. Another prominent example is
the irregular repeat-accumulate (IRA) codes, whose
systematic~\cite{JiKhMc00} and nonsystematic~\cite{PfSaUr05}
versions have both been proved to be capacity-achieving. One
common feature shared by these codes is that they can be
understood to be codes defined on bipartite graphs with variable
nodes and check nodes~\cite{Wi96}, and their iterative decoding
complexity is closely related to the number of edges in their
graphical representations. A fundamental question arises: ``How
simple can the graphs be as a function of their performance?''

In~\cite{SaUr03}, the authors give an information theoretical
lower bound to show that if all variable nodes are transmitted,
then the graphical complexity, \ie, the number of edges per
information bit in the graph, should grow indefinitely as the
multiplicative gap to capacity decreases to 0 on any memoryless
binary-input output-symmetric (MBIOS) channels. This is true even
if maximum-likelihood (ML) decoding is used. On the other hand,
allowing state nodes in the graph, the authors in~\cite{PfSaUr05}
show that nonsystematic IRA codes can achieve capacity on the BEC
with bounded graphical complexity by using the density evolution
method~\cite{RiUr01}. However, partially due to the limitation of
the density evolution method, whether graphs with state nodes can
achieve capacity with bounded graphical complexity on more general
channels other than the BEC still remains unknown.
It should be noted that graphical complexity does not translate
directly to decoding complexity for a general MBIOS channel when
an iterative decoder is utilized. Indeed, it has been conjectured
in~\cite{KhMc01} that for an LDPC code which achieves a fraction
$1-\epsilon$ of the channel capacity, the number of iterations for
achieving vanishing bit error probability grows as $1/\epsilon$
while the average right degree grows as $\ln(1/\epsilon)$, and
thus the average decoding complexity per information bit scales as
$1/\epsilon \ln(1/\epsilon)$. It is thus unclear whether by
reducing the graphical complexity to a constant the conjectured
number of iterations will be influenced. This is an open problem
for general MBIOS channels. Fortunately, at least for the BEC,
this question is resolved, since edges in the graph need only be
visited once when iterative decoding is performed.

In this paper, we introduce a new family of concatenated codes
defined on graphs, namely the concatenated low-density
parity-check and generator matrix (LDPC-GM) codes, and prove that
these codes can achieve capacity using ML decoding on any MBIOS
channels with bounded graphical complexity. These codes are
constructed by serially concatenating an outer LDPC code and an
inner low-density generator matrix (LDGM) code. By deriving and
analyzing the average weight distribution (AWD) and its
corresponding asymptotic growth rate of these codes with a rate-1
LDGM inner code, we show that the inner rate-1 LDGM code can help
eliminate high weight codewords in the LDPC code while maintaining
a vanishing small amount of low weight codewords. The resulting
AWD of these codes thus has an asymptotic growth spectrum, which
can be upper bounded by that of the random ensemble in the
positive region of the curve while the number of codewords in
negative region vanishes at least polynomially in $n$. The ML
performance bound given in~\cite{MiBu01} is then used to prove our
main result. Note that, although the ML performance does not
translate directly to the iterative decoding performance of the
codes, the value of this result is twofold. First, there are
improved iterative decoding algorithms that approach closely the
ML performance~\cite{VaFo04,PiFe04}. Thus, it is conceivable that
the ML performance can be achieved with decoding algorithms having
complexity close to that of iterative decoding. Moreover, this
finding gives a necessary condition for achieving capacity with
suboptimal iterative decoding algorithms without resorting to the
density evolution method, which becomes an infinite dimensional
problem on channels other than the BEC. As a supportive fact on
the potential of these ensembles under iterative decoding, we also
show that these codes can achieve capacity on the BEC with bounded
decoding complexity per information bit for all erasure
probabilities in (0, 1).

The remaining of this paper is organized as follows. We review and
prove some basic properties of the AWD of Gallager's LDPC ensemble
in Section~\ref{LDPC} and derive the average input-output weight
enumerator of the LDGM codes in Section~\ref{LDGM}. Then, in
Section~\ref{concatenation}, we introduce a family of the LDPC-GM
codes and give their AWD and the associated asymptotic growth rate
based on the previous two sections. Detailed analysis on these
LDPC-GM codes is done and the main result is presented in
Section~\ref{analysis}. Allowing the outer LDPC code and inner
LDGM code to be more generally irregular, we prove that the
LDPC-GM codes can achieve capacity on the BEC with bounded
decoding complexity in Section~\ref{DE}. Finally, we conclude this
work in Section~\ref{conclusion}.

\section{The Average Weight Distribution of Gallager's \\ LDPC Ensemble}
\label{LDPC}

Consider Gallager's $(n, j, k)$ LDPC ensemble as introduced
in~\cite{Ga63} with guaranteed rate $R_o = 1-j/k$. Let
$\overline{N_o(l)}$ be the average number of codewords of weight
$l$ in a randomly drawn code from the ensemble. The asymptotic
growth rate of $\overline{N_o(l)}$ is given in~\cite{LiSh02} (it
appears as an upper bound in~\cite{Ga63}) to be
\begin{align} \label{wo}
w_o(a) \triangleq \lim_{n\rightarrow \infty} \frac{1}{n} \ln
\overline{N_o(an)} = \frac{j}{k} \inf_{x>0} \left\{\ln
\frac{(1+x)^k+(1-x)^k}{2 x^{ak}}\right\} -(j-1)H(a)
\end{align}
where $H(a) \triangleq -a\ln a - (1-a)\ln(1-a)$ is the binary
entropy function evaluated with natural logarithms. Some useful
characterizations of $\overline{N_o(l)}$ and $w_o(a)$ are
summarized below.
\begin{fact}\label{fact}
There exists a $\delta_o \in (0, 1/2)$, such that
\begin{enumerate}
\item $\sum_{l = 1}^{n\delta_o} \overline{N_o(l)} = O(n^{-j+2})$.
\item $w_o(a) < 0$ and has exactly one local minimum, but no local maximum  for all $a \in (0, \delta_o)$.
\item $w_o(a) > 0$ for all $a \in (\delta_0, 1/2]$, and $w_o(\delta_o) = 0$.
\item $w_o(a)$ has exactly one local maximum at $a = 1/2$, and $w_o(1/2) = R_o\ln 2$.
\item When $k$ is even, $\overline{N_o(l)} = \overline{N_o(n - l)}$, for
all $l \in \{0, 1, \ldots, n\}$.
\end{enumerate}
\end{fact}
In Fact~\ref{fact}, item 1 to 4 are proved in~\cite[Appendix
A]{Ga63}, and item 5 follows from the linearity of the LDPC codes
and the fact that the all-1 word is always a codeword when $k$ is
even. In order to use item 5, and for other mathematical
convenience, we will assume throughout this paper that $k$ is
even.

We would like to prove two more results, which will help our later
analysis involving LDPC codes. The first lemma gives a close-form
upper bound on $w_o(a)$, which is tight especially when $a$ is
around $1/2$.

\begin{lemma} \label{ubwo}
$w_o(a) \leq (1-R_o)\ln [1 + (1-2a)^k] + [H(a) - (1-R_o)\ln 2]$.
\end{lemma}
\begin{proof}
Bounding the infimum term of~\eqref{wo} by substituting $x =
\frac{a}{1-a}$ proves the lemma.
\end{proof}

The next lemma gives a sufficient condition on $k$ for any desired
lower bound of $\delta_o$, where we denote by $H^{-1}(x)$ the
unique $a \in [0, 1/2]$, such that $H(a) = x$.
\begin{lemma} \label{lbk_do}
Given any $\delta_l \in (0, H^{-1}((1-R_o)\ln 2))$, if
\begin{align}
k > \frac{\ln \left[1-\frac{H(\delta_l)}{(1-R_o)\ln 2}\right]}{\ln
(1-2\delta_l)},
\end{align}
then $\delta_o > \delta_l$.
\end{lemma}
\begin{proof}
After some algebraic manipulations, it can be shown that
\begin{align} \label{lbk_do1}
k > \frac{\ln \left[1-\frac{H(\delta_l)}{(1-R_o)\ln 2}\right]}{\ln
(1-2\delta_l)} \Rightarrow (1-R_o)\ln (1 + (1-2\delta_l)^k) +
[H(\delta_l) - (1-R_o)\ln 2] < 0
\end{align}
Now, the lemma follows from Lemma~\ref{ubwo} and Fact~\ref{fact}.
\end{proof}

\section{The Average Input-Output Weight Enumerator of the LDGM Ensemble}
\label{LDGM}


Consider the regular LDGM ensemble with codeword length $n$ such
that each input node is connected to $c$ check nodes, and each
check node is connected to $d$ input nodes. Let
$\overline{Z_{w,h}}$ be the average number of codewords with input
weight $w$ and output weight $h$ in a code drawn randomly from the
ensemble. We have
\begin{align} \label{aiowe1}
\overline{Z_{w,h}} = \binom{R_i n}{w} P(H=h|W=w),
\end{align}
where $R_i \triangleq d/c$ is the rate of the LDGM codes, and $H$
and $W$ are random variables denoting the input and output weight,
respectively, of a randomly drawn codeword. Now, given the input
weight of the codeword is $w$, the output weight is $h$ if and
only if exactly $h$ check nodes are connected to an odd number of
edges emanated from ``1'' input nodes, and the remaining $n-h$
check nodes are connected to an even number of them. Counting the
number of ways of connecting $cw$ edges to $dn$ check node sockets
such that exactly $h$ check nodes have an odd number of
connections, we see that the value is equal to
\begin{align}
\binom{n}{h}\text{coef}(f_-(x,d)^hf_+(x,d)^{n-h}, x^{cw}),
\end{align}
where $\text{coef}(f(x), x^a)$ denotes the coefficient of $x^a$ in
the polynomial $f(x)$, and $f_-(x,d) \triangleq
\frac{1}{2}[(1+x)^d - (1-x)^d]$ and $f_+(x,d) \triangleq
\frac{1}{2}[(1+x)^d + (1-x)^d]$ are defined to simplify notation.
Since the total number of ways of connecting $cw$ edges to $nd$
sockets is equal to $\binom{nd}{cw}$, we have
\begin{align} \label{aiowe2}
P(H=h|W=w) =
\frac{\binom{n}{h}}{\binom{nd}{cw}}\text{coef}(f_-(x,d)^hf_+(x,d)^{n-h},
x^{cw})
\end{align}
Combining~\eqref{aiowe1} and~\eqref{aiowe2}, we obtain the average
input-output weight enumerator of the $(c, d)$ regular LDGM
ensemble
\begin{align} \label{aiowe}
\overline{Z_{w,h}} = \frac{\binom{nd/c}{w}}{\binom{nd}{cw}}
\binom{n}{h}\text{coef}(f_-(x,d)^hf_+(x,d)^{n-h}, x^{cw}).
\end{align}

\section{Concatenation of LDPC and Rate-1 LDGM Codes}
\label{concatenation}

\begin{figure}[htbp]
  \centering
  \includegraphics[width=\columnwidth]{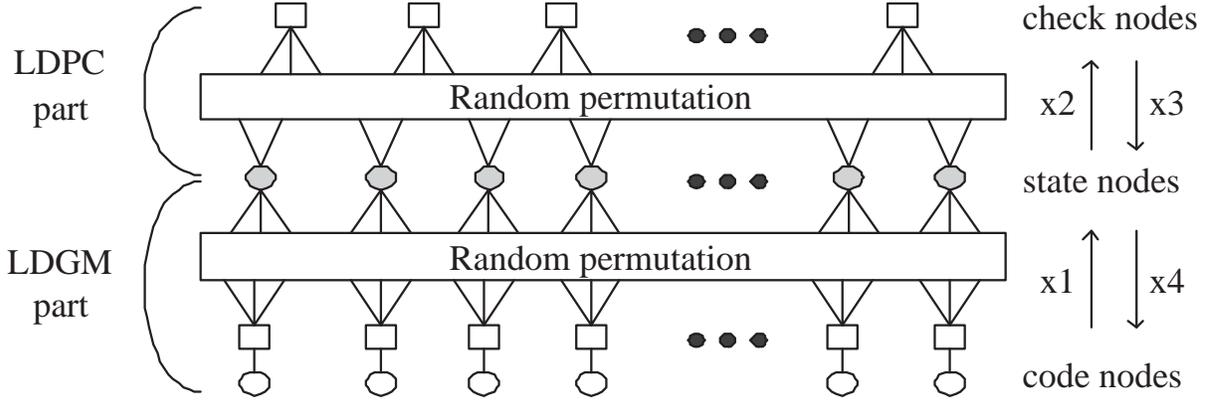}
  \caption{The factor graph of the LDPC-GM codes}
  \label{LDPC-GM_g}
\end{figure}

Consider the concatenation of an outer Gallager's $(n, j, k_1)$
LDPC code and an inner rate-1 $(k_2, k_2)$ regular LDGM code as
shown in Fig.~\ref{LDPC-GM_g}. For simplicity, we assume that $k =
k_1 = k_2$ throughout this paper. If we ignore the possibility
that different LDPC codewords can become the same codeword after
further encoded by the inner LDGM code and just overcount them,
then due to the randomness of the LDPC code construction (although
we do not assume a uniform interleaver between the inner and outer
codes), the AWD of the overall code $\overline{N(l)}$ can be
bounded by
\begin{align} \label{t1}
\overline{N(l)} \leq \overline{N^{ub}(l)} \triangleq \sum_{s =
0}^n \frac{\overline{N_o(s)}\overline{Z_{s,l}}}{\binom{n}{s}} =
\binom{n}{l}\sum_{s = \lceil l/k \rceil}^{\lfloor n-l/k \rfloor}
\frac{\overline{N_o(s)}}{\binom{kn}{ks}}
\text{coef}(f_-(x,k)^lf_+(x,k)^{n-l}, x^{ks}),
\end{align}
where the change of the range of summation in the last equality is
due to the fact that $\text{coef}(f_-(x,k)^lf_+(x,k)^{n-l},
x^{ks}) = 0$ for $s < \lceil l/k \rceil$ and $s > \lfloor n - l/k
\rfloor$. To calculate the asymptotic growth rate of
$\overline{N^{ub}(l)}$, we use the following important equation
given in~\cite{BuMi04}
\begin{align} \label{bur}
&\lim_{\substack{n \rightarrow \infty \\ \text{coef}(f(x), x^{a
n}) \neq 0}} \frac{1}{n}\ln \text{coef}(f(x)^n, x^{a n}) =
\inf_{x>0} \ln \frac{f(x)}{x^a}
\end{align}
where $0 < a < 1$, and $f(x)$ is a polynomial with nonnegative
coefficients. Also used is the well known property of binomial
coefficients
\begin{align} \label{bin}
\lim_{n \rightarrow \infty} \frac{1}{n} \ln \binom{n}{a n} = H(a),
\quad \forall a\in [0, 1]
\end{align}
~\eqref{t1}, ~\eqref{bur} and~\eqref{bin} then give
\begin{align}
w(a) \triangleq& \lim_{n\rightarrow \infty} \frac{1}{n} \ln
\overline{N(an)} \nonumber \\
\leq& \lim_{n\rightarrow \infty} \frac{1}{n} \ln
\overline{N^{ub}(an)} \nonumber \\
=& H(a) + \max_{\frac{a}{k} \leq b \leq 1 - \frac{a}{k}} w_o(b) -
kH(b) + \inf_{x>0} \ln \frac{f_-(x, k)^af_+(x, k)^{1-a}}{x^{bk}}
\nonumber \\
\stackrel{(a)}{\leq}& H(a) + \max_{\frac{a}{k} \leq b \leq 1 -
\frac{a}{k}} w_o(b) + a\ln [1-(1-2b)^k] + (1-a)\ln [1+(1-2b)^k] -
\ln 2 \nonumber
\\
\triangleq& w^{ub}(a)
\end{align}
where (a) follows by substituting $x = \frac{b}{1-b}$ in the
infimum expression. To investigate the true rate $R_1$ of a
randomly drawn LDPC-GM code from this ensemble, let $N(0)$ be the
random variable denoting the number of LDPC codewords which after
encoded by the inner LDGM encoder becomes the all-0 word. Then, we
have by linearity of the LDPC-GM codes and Markov's inequality
that
\begin{align}
P(R_1 < R_o - r) = P(N(0) > 2^{nr}) \leq
\frac{\overline{N(0)}}{2^{nr}} \leq O(2^{n(w^{ub}(0)-r)})
\end{align}
which goes to 0 as $n$ goes to infinity for all $r > w^{ub}(0)$.
Therefore we can define the guaranteed rate of these LDPC-GM codes
with asymptotically high probability to be
\begin{align} \label{defrate}
R \triangleq R_o - \max\{w^{ub}(0), 0\}
\end{align}

\section{Analysis of the LDPC-GM Codes}
\label{analysis}

In this section, we will first characterize $w^{ub}(a)$, and then
use the derived results to prove that LDPC-GM codes can be
capacity-achieving on the MBIOS channels using ML decoding with
bounded graphical complexity. Although $w^{ub}(a)$ is not
symmetric about $a = 1/2$, the following lemma shows that we can
focus on analyzing $w^{ub}(a)$ for $a \in [0, 1/2]$ and bound
$w^{ub}(a)$ by $w^{ub}(1-a)$ for $a \in [1/2, 1]$.
\begin{lemma} \label{half}
$w^{ub}(a) \leq w^{ub}(1-a)$ for all $a \in [0, 1/2]$.
\end{lemma}
\begin{proof}
It follows from the facts that
\begin{align} \label{bigsmall}
\ln[1 - (1 - 2b)^k] \leq 0 \leq \ln[1 + (1 - 2b)^k] \quad \forall
b \in [0, 1]
\end{align}
and $a \leq 1-a$ for all $a \in [0, 1/2]$.
\end{proof}

In the next theorem, we prove that given any $R'$ in [0, 1], the
positive part of $w^{ub}(a)$ can be upper bounded by $H(a) -
(1-R_o)\ln 2$ if $k$ is sufficiently large for all $R_o \in [0,
R']$. In this case, we also prove that $\overline{N^{ub}(l)}$ at
least decreases polynomially with $n$ in the negative part of
$w^{ub}(a)$ when $j \geq 3$.
\begin{theorem}\label{wcon}
For any $R' \in [0, 1]$, there exists an integer $M < \infty$ such
that for all $k > M$ and $R_o \in [0, R']$, there exists a
$\delta' < H^{-1}((1-R_o)\ln 2)$ such that the following two
things are true.
\begin{enumerate}
\item
\begin{align}
w^{ub}(a)
\begin{cases}
\leq 0& \text{if $a = 0$}, \\
< 0& \text{if $a \in (0, \delta']$}, \\
\leq H(a) - (1-R_o)\ln 2& \text{if $a \in (\delta', 1/2]$}.
\end{cases}
\end{align}
\item $\overline{N^{ub}(l)} = O(n^{-j+2})$ for all $l \in (0,
\delta' n] \cup [n - \delta' n, n]$.
\end{enumerate}
\end{theorem}
\begin{proof}
See Appendix~\ref{wcon_pf}
\end{proof}

From the above theorem and the definition of the guaranteed rate
of the LDPC-GM codes in~\eqref{defrate}, we have the following
corollary, which says that the conditions implied by the above
theorem also guarantee no rate reduction for the LDPC-GM codes.
\begin{corollary} \label{rate}
If $k > M$, where $M$ is as defined in Theorem~\ref{wcon} for some
$R' \in [0, 1]$, then $R = R_o$ for all $R_o \in [0, R']$.
\end{corollary}

Moreover, if we let $d_{min}$ and $N(l)$ be the random variables
denoting the minimum distance and number of codewords of weight
$l$, respectively, of a randomly drawn code from the concatenated
ensemble, and let $\delta_{GV} = H^{-1}((1-R)\ln 2)$ be the
normalized Gilbert-Varshamov distance, then from Markov's
inequality and Theorem~\ref{wcon}, we have
\begin{align}
P(d_{min} < \delta_{GV} n) =& P\left(\sum_{l \in (0, \delta_{GV}
n)}
N(l) \geq 1\right) \nonumber \\
\leq& \sum_{l \in (0, \delta_{GV} n)} \overline{N(l)} \nonumber \\
\leq& n \max_{l \in (0, \delta'n]} \overline{N^{ub}(l)}  + n
\exp\{n \max_{a \in (\delta',
\delta_{GV})} w^{ub}(a) + o(n)\} \nonumber \\
=& O(n^{-j+3})
\end{align}
which goes to 0 asymptotically as $n$ goes to infinity when $j
\geq 4$. Therefore, we have the following corollary.

\begin{corollary}
If $k > M$, where $M$ is as defined in Theorem~\ref{wcon} for some
rate $R$ and $j \geq 4$, then the LDPC-GM codes have a normalized
minimum distance greater than or equal to the Gilbert-Varshamov
bound with asymptotically high probability.
\end{corollary}

\begin{figure}[htbp]
  \centering
  \includegraphics[width=0.8\columnwidth]{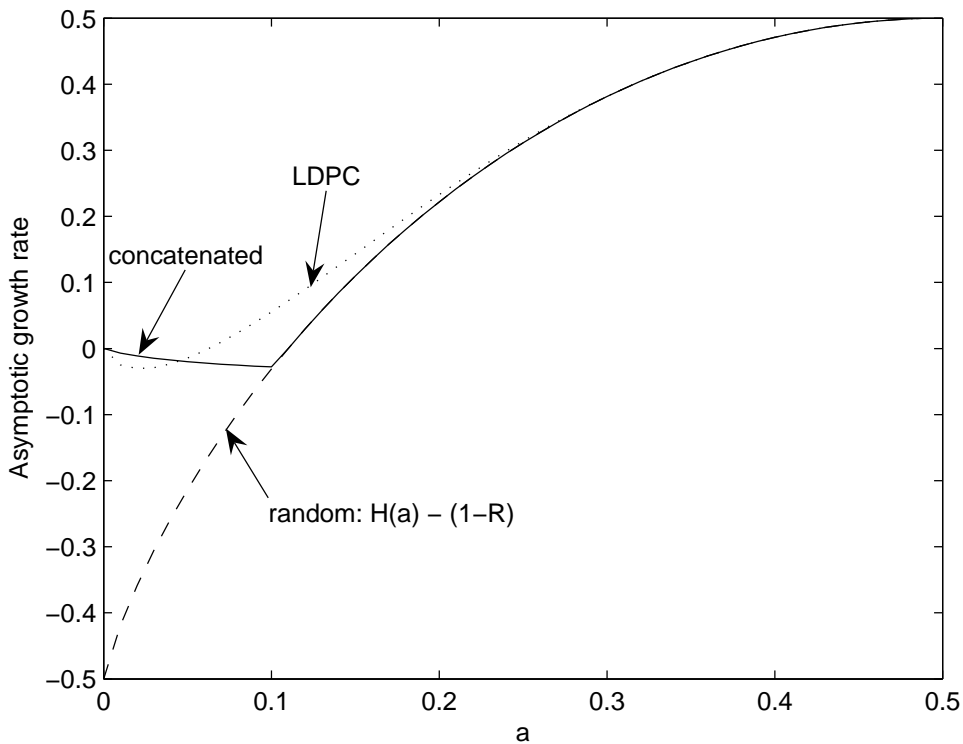}
  \caption{Comparison of $w_o(a)$, $w^{ub}(a)$ and $H(a) - (1-R)$ with $R = 0.5$ and $k = 8$. The logarithm is to the base 2 in
this figure.}
  \label{comagr}
\end{figure}

In Fig.~\ref{comagr}, we compare the asymptotic growth rate of the
LDPC, the LDPC-GM and the random ensemble with $R = 0.5$ and $k =
8$. It is evident that the rate-1 LDGM inner code really helps
eliminate high weight codewords in the outer LDPC code. As a
trade-off, the growth rate of some low weight codewords increases
slightly. However, as long as the growth rate of the low weight
codewords remains negative, Theorem~\ref{wcon} shows that they
still vanish as $n$ goes to infinity.

We are now ready to state our main theorem, which shows that given
any MBIOS channel, there always exists a finite value $M$ such
that these LDPC-GM codes with $k > M$ is capacity-achieving.
\begin{theorem} \label{main}
Given any MBIOS channel with capacity $C$, there exists an integer
$M < \infty$ such that if $k > M$, then $R = R_o$ for the LDPC-GM
ensembles with $R_o < C$. Moreover, for the given channel, the
average block error probability $P_B$ of the LDPC-GM ensembles
with $k > M$, $j \geq 4$ and $R < C$ is vanishingly small when ML
decoding is used.
\end{theorem}
\begin{proof}
See Appendix~\ref{main_pf}.
\end{proof}

As can be seen in Fig.~\ref{LDPC-GM_g}, the graphical complexity
$\Delta$ of these LDPC-GM codes can be evaluated as follows.
\begin{align}
\Delta = \frac{n(j+k) + n}{Rn} = \frac{(2-R)k + 1}{R}
\end{align}
Since Theorem~\ref{main} says that $k$ need not go to infinity to
achieve capacity, we can deduce that these LDPC-GM codes with any
rate $R \in (0, 1)$ can be capacity achieving with bounded
graphical complexity.

\section{Density Evolution for LDPC-GM Codes on the BEC}
\label{DE}

Although the aforementioned LDPC-GM ensembles have finite
graphical complexity, the decoding complexity under ML decoding is
still exponential. In this section, we show that by allowing the
outer LDPC codes to be more generally irregular, the LDPC-GM
ensemble can be capacity-achieving on the BEC under BP decoding
with bounded decoding complexity per information bit. Although
this is not a proof that the same might be true for the MBIOS
channels, it is a good indication of the potential of the LDPC-GM
codes.

Consider the concatenation of a $(\lambda, \rho)$ irregular LDPC
code and a $(2,2)$ regular LDGM code, where $\lambda$ and $\rho$
are the standard variable and check node degree distributions,
respectively, from the edge perspective as defined
in~\cite{RiShUr01}. Note that this LDPC-GM ensemble has guaranteed
rate
\begin{align}
R = 1 - \frac{\int_0^1 \rho(t)dt}{\int_0^1 \lambda(t)dt}
\end{align}
and our task is to successfully decode the non-transmitted LDPC
codewords. Let $q$ be the channel erasure probability, and let
$x_1$, $x_2$, $x_3$ and $x_4$ be the probabilities of erasure on
edges from check to variable(LDGM), variable to check(LDPC), check
to variable(LDPC) and variable to check(LDGM), respectively as
shown in Fig.~\ref{LDPC-GM_g}. Then, assuming we are operating at
some fixed point, we have the following density evolution
equations.
\begin{subequations} \label{de_LDPC-GM}
\begin{align}
x_1 =& 1 - (1-q)(1-x_4) \\
x_2 =& x_1^2\lambda(x_3) \\
x_3 =& 1 - \rho(1-x_2) \\
x_4 =& x_1 \tilde{\lambda}(x_3)
\end{align}
\end{subequations}
where $\tilde{\lambda}(x) = \sum_{i=1}^\infty
\tilde{\lambda}_ix^i$ and
\begin{align} \label{variabledis_node}
\tilde{\lambda}_i = \frac{\lambda_i/i}{\int_0^1\lambda(t)dt},
\end{align}
which denotes the fraction of variable nodes in the LDPC code with
degree $i$. Equivalently, we have
\begin{align}
\tilde{\lambda}(x) =
\frac{\int_0^x\lambda(t)dt}{\int_0^1\lambda(t)dt}
\end{align}
Note that equations~\eqref{de_LDPC-GM} are also the density
evolution equations for the serially concatenated codes with an
outer LDPC code and an inner differentiator code. Solving these
equations for $x_3$, we have
\begin{align} \label{fixp_LDPC-GM}
x_3 =
\rho\left(1-\left[\frac{q}{1-(1-q)\tilde{\lambda}(x_3)}\right]^2\lambda(x_3)\right)
\end{align}
If~\eqref{fixp_LDPC-GM} has no solution in $(0, 1]$, then $x_3$
must converge to 0 and thus $x_4$ must converge to 0 as the number
of iterations goes to infinity. Therefore, if we have
\begin{align} \label{desuff_LDPC-GM}
1 -
\rho\left(1-\left[\frac{q}{1-(1-q)\tilde{\lambda}(x_3)}\right]^2\lambda(x_3)\right)
< x_3, \quad \forall x_3 \in (0, 1]
\end{align}
then the BP decoding is successful. Note that~\eqref{fixp_LDPC-GM}
is essentially the same as equation (6) in~\cite{PfSaUr05} except
for the following changes: $x_0 \rightarrow 1 - x_3$, $p
\rightarrow 1-q$, $\lambda(\cdot) \rightarrow \rho(\cdot)$,
$\rho(\cdot) \rightarrow \lambda(\cdot)$, and $R(\cdot)
\rightarrow \tilde{\lambda}(\cdot)$. More
generally,~\eqref{fixp_LDPC-GM} is an instance of the symmetry
introduced in~\cite{PfSa05}. So, in the following, we will use the
results proved in~\cite{PfSaUr05} to show two particular degree
distribution pairs are capacity-achieving under BP decoding.
\begin{theorem}[Check-regular ensemble] \label{cregular}
Let
\begin{align}
\lambda(x) =& \frac{1 - (1-x)^\frac{1}{k-1}}{\left[1 -
(1-q)\left(1-kx+(k-1)\left[1-(1-x)^{\frac{k}{k-1}}\right]\right)\right]^2} \\
\rho(x) =& x^{k-1}
\end{align}
Then for $k = 3$ and $q \in [\frac{12}{13}, 1)$, $\lambda(x)$ has
only non-negative coefficients. Moreover, for any $\epsilon \in
(0,1)$, let $M(\epsilon)$ be the smallest positive integer such
that\footnote{$M(\epsilon)$ exists for all $\epsilon \in (0, 1)$
since $\sum_{i = 1}^\infty \frac{\lambda_i}{i} = \int_0^1
\lambda(t)dt = \frac{1}{qk}$, which means $\sum_{i = M(\epsilon) +
1}^\infty \frac{\lambda_i}{i}$ can be made arbitrarily close to 0
by increasing $M(\epsilon)$.}
\begin{align}
\sum_{i = M(\epsilon) + 1}^\infty \frac{\lambda_i}{i} <
\frac{\epsilon(1-q)}{qk}
\end{align}
and let $\lambda_\epsilon(x)$ be the truncated degree distribution
of $\lambda(x)$ by treating all variable nodes with degree greater
than $M(\epsilon)$ as pilot bits. Then the degree distribution
pair $(\lambda_\epsilon, \rho)$ achieves a fraction $1-\epsilon$
of the channel capacity with vanishing bit error probability under
BP decoding.
\end{theorem}
\begin{proof}
See Appendix~\ref{cregular_pf}.
\end{proof}
The decoding complexity per information bit of this check-regular
ensemble can be calculated as follows
\begin{align}
\Delta < \frac{knq + 2n + n}{(1-q)(1-\epsilon)n} = \frac{qk +
3}{(1-q)(1-\epsilon)},
\end{align}
which approaches the bounded value $\frac{qk + 3}{1-q}$ as
$\epsilon$ goes to 0.

\begin{theorem}[Variable-regular ensemble] \label{vregular}
Let
\begin{align}
\lambda(x) =& x^2\\
\rho(x) =& 1 + \frac{2(1-q)(1-x)^2
\sin\left(\frac{1}{3}\arcsin\left(\sqrt{-\frac{27(1-q)(1-x)^{\frac{3}{2}}}{4q^3}}\right)\right)}{\sqrt{3}q^4\left[-\frac{(1-q)(1-x)^{\frac{3}{2}}}{q^3}\right]^\frac{3}{2}}
\end{align}
Then for $q \in [0.05, 1]$, $\rho(x)$ has only non-negative
coefficients. Moreover, for any $\epsilon \in (0,1)$, let
$M(\epsilon)$ be the smallest positive integer such
that\footnote{$M(\epsilon)$ exists for all $\epsilon \in (0, 1)$
since $\sum_{i = 1}^\infty \rho_i = 1$, which means $\sum_{i =
M(\epsilon) + 1}^\infty \rho_i$ can be made arbitrarily close to 0
by increasing $M(\epsilon)$.}
\begin{align}
\sum_{i = M(\epsilon) + 1}^\infty \rho_i < \frac{\epsilon(1-q)}{3}
\end{align}
and let
\begin{align}
\rho_\epsilon(x) \triangleq \left(1 - \sum_{i = 1}^{M(\epsilon)}
\rho_i\right) + \sum_{i = 1}^{M(\epsilon)} \rho_i x^{i-1}
\end{align}
be the truncated degree distribution of $\rho(x)$. Then the degree
distribution pair $(\lambda, \rho_\epsilon)$ achieves a fraction
$1-\epsilon$ of the channel capacity with vanishing bit error
probability under BP decoding.
\end{theorem}
\begin{proof}
See Appendix~\ref{vregular_pf}.
\end{proof}
The decoding complexity per information bit of this
variable-regular ensemble can be calculated as follows
\begin{align}
\Delta < \frac{3n + 2n + n}{(1-q)(1-\epsilon)n} =
\frac{6}{(1-q)(1-\epsilon)}
\end{align}
which approaches the bounded value $\frac{6}{1-q}$ as $\epsilon$
goes to 0.

One drawback of the these capacity-achieving degree distribution
pairs is that they are not guaranteed to be valid, \ie, with only
nonnegative coefficients, for all $q \in (0, 1)$. However, since
they are valid for $q$ close 1 (which is not true for the
capacity-achieving IRA codes in~\cite{PfSaUr05}), this problem can
be solved by considering punctured LDPC-GM codes.
In~\cite{PiRaFe05}, it is shown that random puncturing results in
no performance loss on the gap to capacity for codes on the BEC.
Hence, it follows that puncturing can be used to increase the rate
of the LDPC-GM codes without affecting its capacity-achievability,
a fact that was also observed by Pfister and Sason~\cite{PfSa05}.
Furthermore, since a punctured LDPC-GM ensemble can also be viewed
as another unpunctured LDPC-GM ensemble with inner irregular LDGM
codes (which is no longer rate-1 in general), we have the
following theorem.
\begin{theorem} \label{pldpc-gm}
Let $(\lambda, \rho)$ be a degree distribution pair implied by
Theorem~\ref{cregular} or Theorem~\ref{vregular} for some given
$\epsilon$ and $q'$. Consider the LDPC-GM ensemble, whose outer
LDPC code has degree distribution pair $(\lambda, \rho)$ from the
edge perspective, and inner LDGM code has degree distribution pair
$(F, G)$ from the node perspective\footnote{That is, $F(x) =
\sum_{i=0}^\infty F_ix^i$ and $G(x) = \sum_{i=1}^\infty G_ix^i$,
where $F_i$ and $G_i$ denote the fraction of input and check nodes
that have $i$ neighboring check and input nodes, respectively, in
the LDGM code.}. Then for any given $p \in [0, q']$, if
\begin{align}
F(x) =& [x(1-p) + p]^2 \\
G(x) =& x^2
\end{align}
then this LDPC-GM ensemble achieves a fraction of $1-\epsilon$ of
the channel capacity on the BEC with erasure probability $q
\triangleq \frac{q'-p}{1-p}$ under BP decoding.
\end{theorem}
\begin{proof}
See Appendix~\ref{pldpc-gm_pf}.
\end{proof}
This theorem says that, given any capacity-achieving degree
distribution pair $(\lambda, \rho)$ for some erasure probability
$q'$, we can generate capacity-achieving LDPC-GM ensembles for all
erasure probabilities $q \in [0, q']$ by adjusting $p$. Since $q'$
can be arbitrarily close to 1, and the maximum degrees of $F$ and
$G$ are bounded for all $p$, this construction can be done to
produce capacity-achieving LDPC-GM ensembles for all rate in (0,
1) on the BEC with bounded decoding complexity.

\section{Conclusion}
\label{conclusion}

In this paper, the LDPC-GM codes, \ie, the concatenated codes with
an outer LDPC code and an inner LDGM code, are introduced. In the
case that the outer code is Gallager's $(n, j, k)$ LDPC code and
the inner code is a rate-1 $(k, k)$ regular LDGM code, we prove
that for any desired range of rates $R_o$, there always exists an
integer $M < \infty$ such that if $k > M$ then the inner LDGM
encoder results in no rate reduction for the outer LDPC code.
Moreover, the LDGM encoder helps eliminate high weight codewords
while maintaining a vanishingly small amount of low weight
codewords in the LDPC code. The resulting asymptotic growth
spectrum of the LDPC-GM codes has a positive part, which can be
upper bounded by the asymptotic growth spectrum of the random
ensemble, and a negative part, where the number of codewords
vanishes at least polynomially in $n$ when $j \geq 4$. Note that,
the condition $j \geq 4$ is automatically satisfied when $k$ is
big enough. It then follows easily that these codes achieve the
Gilbert-Varshamov bound with asymptotically high probability.
Furthermore, after applying the ML performance bound given
in~\cite{MiBu01} to these LDPC-GM codes, we prove that they can
achieve capacity on any MBIOS channels using ML decoding. Since
all these results are implied by the only condition that $k$ is
greater than some finite number, which shows that the number of
edges per information bit in the graph need not go to infinity to
achieve capacity, we have proved that these LDPC-GM codes are
capacity-achieving codes with bounded graphical complexity on any
MBIOS channels.

On the other hand, if the outer LDPC code is allowed to be
irregular, then invoking the density evolution method, we use the
results in~\cite{PfSaUr05} to show two particular ensembles of the
LDPC-GM codes can achieve capacity on the BEC under BP decoding
with bounded decoding complexity. Moreover, extensions valid for
all erasure probabilities of the BEC using inner irregular LDGM
codes are also presented. These favorable results could suggest
high potential of the LDPC-GM codes to achieve capacity on the
MBIOS channels with bounded decoding complexity per iteration.
However, since the scaling on the required number of iterations
for successful iterative decoding for the LDPC-GM codes compared
to other existing codes remains unknown, whether this bounded
graphical complexity property implies bounded decoding complexity
using iterative decoding is still an open problem on MBIOS
channels.

\appendix
\section{Proof of Theorem~\protect\ref{wcon}}
\label{wcon_pf}

\begin{enumerate}
\item Define
\begin{align}
f(b) \triangleq w_o(b) + a\ln \frac{1-(1-2b)^k}{2} + (1-a)\ln
\frac{1+(1-2b)^k}{2}
\end{align}
We will bound $f(b)$ in two cases. By Lemma~\ref{lbk_do}, for any
$\delta_l \in (0, H^{-1}((1-R')\ln 2)) \subset (0,
H^{-1}((1-R_o)\ln 2))$, if
\begin{align}
k > M_1 \triangleq \frac{\ln \left[1-\frac{H(\delta_l)}{(1-R')\ln
2}\right]}{\ln (1-2\delta_l)} \geq \frac{\ln
\left[1-\frac{H(\delta_l)}{(1-R_o)\ln 2}\right]}{\ln
(1-2\delta_l)},
\end{align}
then $w_o(\delta_l) < 0$ for all $R_o \in [0, R']$. Therefore, for
$k > M_1$ and $b \in [a/k, \delta_l] \cup [1-\delta_l, 1-a/k]$ (we
assume without loss of generality that $a/k \leq \delta_l$.
otherwise, we just skip this step), we have
\begin{align}
f(b) \leq w_o(b) - H(a) \leq \max\{w_o(a/k), w_o(\delta_l)\} -
H(a),
\end{align}
where the first inequality follows from the fact that relative
entropy is always nonnegative, and the second inequality follows
from Fact~\ref{fact}. On the other hand, when $b \in (\delta_l,
1-\delta_l)$, we have from Lemma~\ref{ubwo} that
\begin{align}
f(b) \leq& (1-R_o)\ln [1 + (1-2b)^k] + H(b) - (1-R_o)\ln 2 +
\nonumber \\
&+a\ln\frac{1-(1-2b)^k}{2} + (1-a)\ln \frac{1+(1-2b)^k}{2} \nonumber \\
\leq& -(1-R_o)\ln 2 - \ln 2 + \{H(b) + (2-R_o-a)\ln [1+(1-2b)^k]\}
\nonumber \\
\leq& -(1-R_o)\ln 2 - \ln 2 + \{H(b) + 2\ln [1+(1-2b)^k]\}
\end{align}
where the last two inequalities follow from~\eqref{bigsmall}.
Since
\begin{align}
\frac{\partial^2 H(b)}{\partial b^2} = -\frac{1}{(1-b)b} \leq -4,
\end{align}
and
\begin{align}
\frac{\partial^2 2\ln [1+(1-2b)^k]}{\partial b^2} =&
\frac{8k[k - 1 - (1-2b)^k](1-2b)^{k-2}}{[1+(1-2b)^k]^2} \nonumber \\
\leq& 8k(k-1)(1-2b)^{k-2} \nonumber \\
\leq& 8k(k-1)(1-2\delta_l)^{k-2},
\end{align}
which can be made arbitrarily close to 0 for a large enough $k$,
there exists a $M_2$ such that $k
> M_2$ implies that the maximum of $H(b) + 2 \ln
[1+(1-2b)^k]$ is attained at $b = 1/2$, and thus
\begin{align} \label{t2}
f(b) \leq -(1-R_o)\ln 2, \quad \forall b \in (\delta_l,
1-\delta_l)
\end{align}
Therefore, for all $k > M \triangleq \max\{M_1, M_2\}$, we have
\begin{align}
w^{ub}(a) = H(a) + \max_{\frac{a}{k} \leq b \leq 1 - \frac{a}{k}}
f(b) \leq \max\{H(a) - (1-R_o)\ln 2, w_o(a/d), w_o(\delta_l)\}
\end{align}
Since $\max\{w_o(a/d), w_o(\delta_l)\} < 0$ for all $a > 0$, there
must exist a $\delta' < H^{-1}((1-R_o)\ln 2)$ such that this part
of the theorem is true.

\item For all $l \in (0, \delta' n] \cup [n - \delta' n, n]$ and $k > M$, we have
\begin{align} \label{combin}
\overline{N_c^{ub}(l)}=& \sum_{s = \lceil l/k \rceil}^{\lfloor
n-l/k \rfloor}
\frac{\overline{N_o(s)}\overline{Z_{s,l}^{(LDPG)}}}{\binom{n}{s}}
\nonumber \\
\stackrel{(a)}{\leq}& \sum_{s = \lceil l/k \rceil}^{\delta_l n}
\overline{N_o(s)} + \sum_{s = n - \delta_l n}^{\lfloor n-l/k
\rfloor} \overline{N_o(s)} + \sum_{s = \delta_l n}^{n - \delta_l
n}
\frac{\overline{N_o(s)}\overline{Z_{s,l}^{(LDPG)}}}{\binom{n}{s}}
\nonumber \\
\stackrel{(b)}{\leq}& O(n^{-j+2}) + n \exp\{n [H(l/n) +
\max_{\delta_l \leq b \leq
1-\delta_l}f(b)] + o(n)\} \nonumber \\
\stackrel{(c)}{\leq}& O(n^{-j+2}) + n \exp\{n [H(l/n) - (1-R)\ln 2] + o(n)\} \nonumber \\
\stackrel{(d)}{=}& O(n^{-j+2})
\end{align}
where $o(n)$ denotes some value that converges to 0 as $n$ goes to
infinity. In~\eqref{combin}, (a) follows from the fact that
$\overline{Z_{s,l}^{(LDPG)}} / \binom{n}{s} \leq 1$ since it is a
probability as shown in~\eqref{aiowe1}, (b) follows from
Fact~\ref{fact}, (c) follows from~\eqref{t2}, and (d) follows from
the fact that $\delta' < H^{-1}((1-R)\ln 2)$.
\end{enumerate}

\section{Proof of Theorem~\protect\ref{main}}
\label{main_pf}

Let $M$ be as defined in Theorem~\ref{wcon} for $R' = C$. Then by
Corollary~\ref{rate}, we have $R = R_o$ for all $k > M$ and $R_o <
C$. Let $U \subset \{1, 2, \ldots, n\}$, and $U^c$ be its
complementary set. The following upper bound on the average block
error probability under ML decoding is given in~\cite{MiBu01}
\begin{align} \label{ufs0}
P_B \leq \sum_{l \in U} \{\overline{N(l)}D^l\} + 2^{-n E_r(R +
\frac{\ln \alpha}{n \ln 2})}
\end{align}
where
\begin{align}
\alpha \triangleq \max_{l \in U^c} \frac{\overline{N(l)}}{2^{nR} -
1}\frac{2^{n}}{\binom{n}{l}}
\end{align}
$E_r(\cdot)$ is the random coding exponent, and
\begin{align} \label{D}
D \triangleq \sum_y \sqrt{p(y|0)p(y|1)} \leq 1
\end{align}
where $p(y|0)$ and $p(y|1)$ are the conditional probability
density functions of the output of the MBIOS channel given the
input. If we apply this bound to the LDPC-GM ensemble with $k
> M$ and $R_o < C$, and let
\begin{align}
U \triangleq \left\{l: \frac{l}{n} \in (0, \delta'] \cup
[1-\delta', 1]\right\},
\end{align}
where $\delta'$ is as defined in Theorem~\ref{wcon}, then we have
from Theorem~\ref{wcon} that
\begin{align} \label{ufs1}
\sum_{l \in U} \{\overline{N(l)}D^l\} \leq \sum_{l \in U}
\overline{N^{ub}(l)} \leq n O(n^{-j+2}) = O(n^{-j+3})
\end{align}
But we have from the same theorem and Lemma~\ref{half} that
\begin{align} \label{ufs2}
\lim_{n \rightarrow \infty} \frac{\ln \alpha}{n} =& \max_{a \in
(\delta', 1-\delta')} w(a) - [H(a) - (1-R)\ln 2]
\nonumber \\
\leq& \max_{a \in
(\delta', 1/2]} w^{ub}(a) - [H(a) - (1-R)\ln 2] \nonumber \\
\leq& 0
\end{align}
Hence we have
\begin{align}
P_B \leq O(n^{-j+3}) + 2^{-n E_r(R)}
\end{align}
which goes to 0 as $n$ goes to infinity for all $R < C$ and $j
\geq 4$. Thus, the theorem is proved.

\section{Proofs of Section~\protect\ref{DE}} \label{DE_pf}

First, we need a lemma.
\begin{lemma} \label{de_rate_LDPC-GM}
If the degree distribution pair $(\lambda, \rho)$ satisfies
$\rho(0) = 0$, $\rho(1) = 1$, and satisfies~\eqref{fixp_LDPC-GM}
for all $x_3 \in [0, 1]$, then $R = 1-q$.
\end{lemma}
\begin{proof}
\cite[Lemma 1]{PfSaUr05} shows that under the assumed conditions,
we have
\begin{align}
\frac{\int_0^1 \rho(t)dt}{\int_0^1 \lambda(t)dt} = q
\end{align}
\end{proof}

\subsection{Proof of Theorem~\protect\ref{cregular}} \label{cregular_pf}

The facts that $(\lambda, \rho)$ satisfies~\eqref{fixp_LDPC-GM}
for all $x \in [0, 1]$ and that $\lambda(x)$ has only non-negative
coefficients for $k = 3$ and $q \in [\frac{12}{13}, 1)$ are proved
in~\cite[Theorem 1]{PfSaUr05}. By the definition of
$\lambda_\epsilon$, we have effectively
\begin{align}
\lambda_\epsilon (x) = \sum_{i = 1}^{M(\epsilon)}\lambda_i x^{i-1}
\end{align}
in the density evolution equations. Hence, it follows that
$\lambda_\epsilon (x) < \lambda(x)$, and the corresponding
$\tilde{\lambda}_\epsilon (x) < \tilde{\lambda}(x)$ for all $x \in
(0, 1]$. Therefore, ~\eqref{desuff_LDPC-GM} is satisfied, which
implies that the BP decoding is successful. To find the rate of
this ensemble of codes, let
\begin{align}
\delta \triangleq \sum_{M(\epsilon)+1}^\infty \tilde{\lambda}_i
\end{align}
be the fraction of pilot nodes. Then, we have
\begin{align}
R =& \frac{(1-\delta)\int_0^1 \lambda(t)dt - \int_0^1
\rho(t)dt}{\int_0^1 \lambda(t)dt} \nonumber \\
=& 1 - \delta - \frac{\int_0^1 \rho(t)dt}{\int_0^1 \lambda(t)dt}
\nonumber \\
=& 1 - q - \delta
\end{align}
where the last equality follows from the facts that $\rho(0) = 0$,
$\rho(1) = 1$, and Lemma~\ref{de_rate_LDPC-GM}. But,
from~\eqref{variabledis_node}
\begin{align}
\delta = \sum_{M(\epsilon)+1}^\infty
\frac{\lambda_i/i}{\int_0^1\lambda(t)dt}  =
q\sum_{M(\epsilon)+1}^\infty \frac{\lambda_i/i}{\int_0^1\rho(t)dt}
= qk\sum_{M(\epsilon)+1}^\infty \lambda_i/i  < \epsilon(1-q)
\end{align}
Therefore, it follows that $R > (1-\epsilon)(1-q)$, and the
theorem is proved.

\subsection{Proof of Theorem~\protect\ref{vregular}} \label{vregular_pf}

The facts that $(\lambda, \rho)$ satisfies~\eqref{fixp_LDPC-GM}
for all $x \in [0, 1]$ and that $\rho(x)$ has only non-negative
coefficients for $q \in [0.05, 1]$ are proved in~\cite[Theorem
2]{PfSaUr05}. Since $\rho_\epsilon(x) > \rho(x)$ for all $x \in
(0, 1]$, ~\eqref{desuff_LDPC-GM} is satisfied and the BP decoding
is successful. As for the rate of this ensemble of codes, we have
\begin{align}
R =& 1 - \frac{\int_0^1
\rho_\epsilon(t)dt}{\int_0^1 \lambda(t)dt} \nonumber \\
=& 1 - \frac{\sum_{i = 1}^{M(\epsilon)} \frac{\rho_i}{i} + 1 -
\sum_{i = 1}^{M(\epsilon)}\rho_i}{\int_0^1 \lambda(t)dt} \nonumber \\
>& 1 - \frac{\sum_{i = 1}^{\infty} \frac{\rho_i}{i} + 1 - \sum_{i
= 1}^{M(\epsilon)}\rho_i}{\int_0^1 \lambda(t)dt} \nonumber \\
=& 1 - \frac{\int_0^1 \rho(t)dt + \sum_{i
= M(\epsilon) + 1}^\infty \rho_i}{\int_0^1 \lambda(t)dt} \nonumber \\
\stackrel{(a)}{=}& 1 - q - 3 \sum_{i = M(\epsilon) + 1}^\infty \rho_i \nonumber \\
>& (1-\epsilon)(1-q)
\end{align}
where (a) follows from the facts that $\rho(0) = 0$, $\rho(1) =
1$, and Lemma~\ref{de_rate_LDPC-GM}. Hence, the theorem is proved.

\subsection{Proof of Theorem~\protect\ref{pldpc-gm}} \label{pldpc-gm_pf}

Let $f$ be the degree distribution corresponding to $F$ from the
edge perspective. We have
\begin{align}
f(x) = \frac{F'(x)}{F'(1)} = x(1-p) + p
\end{align}
and the following set of density evolution equations
\begin{subequations} \label{de_LDPC-GMi}
\begin{align}
x_1 =& 1 - (1-q')(1-x_4) \\
x_2 =& F(x_1)\lambda(x_3) \\
x_3 =& 1 - \rho(1-x_2) \\
x_4 =& f(x_1) \tilde{\lambda}(x_3)
\end{align}
\end{subequations}
After some algebraic manipulations, the fixed point equation can
be shown to be
\begin{align}
x_3 = 1 - \rho\left(1-\left[\frac{q'(1-p) +
p}{1-(1-q')(1-p)\tilde{\lambda}(x_3)}\right]^2\lambda(x_3)\right)
\end{align}
which is the same as~\eqref{fixp_LDPC-GM} if we let $q$ be as
defined in this theorem. Hence, from Theorem~\ref{cregular} and
Theorem~\ref{vregular}, the decoding is successful under BP
decoding on the BEC with erasure probability $q$. Moreover, the
rate of this ensemble is given by
\begin{align}
R =& \{\text{rate of the outer LDPC code}\} \times
\frac{\{\text{number of input nodes in the LDGM code}\}}{\{\text{number of check nodes in the LDGM code}\}} \nonumber \\
=& \{\text{rate of the outer LDPC code}\} \times
\frac{G'(1)}{F'(1)} \nonumber \\
>& (1-\epsilon) (1-q')\frac{1}{1-p} \nonumber
\\
=& (1-\epsilon)(1-q),
\end{align}
which then proves this theorem.

\bibliographystyle{IEEEbib}
\bibliography{Hsuref}

\end{document}